\documentclass[11pt]{nature_mod}
\usepackage{url}
\usepackage{amsmath}
\usepackage{amssymb}
\usepackage{amstext}
\usepackage{bigstrut}
\usepackage{bm}
\usepackage{booktabs}
\usepackage[bf,footnotesize]{caption}
\usepackage{chngpage}
\usepackage{color}
\usepackage{dcolumn}
\usepackage{float}
\usepackage{graphicx}
\usepackage{lineno}
\usepackage{makecell}
\usepackage{mathrsfs}
\usepackage{multirow}
\usepackage{rotating}
\usepackage{subfigure}
\usepackage{threeparttable}
\usepackage{xr}
\usepackage{algorithm}
\usepackage{algpseudocode}

\definecolor{darkred}{rgb}{0.75,0,0}
\definecolor{darkgreen}{rgb}{0,0.5,0}
\definecolor{darkblue}{rgb}{0,0,0.75}
\definecolor{darkorange}{rgb}{1,0.9,0.1}
\definecolor{dark}{rgb}{0,0,0}

\DeclareMathAlphabet{\mcal}{OMS}{cmsy}{m}{n}  

\usepackage{caption}
\newcommand*\patchAmsMathEnvironmentForLineno[1]{%
	\expandafter\let\csname old#1\expandafter\endcsname\csname #1\endcsname
	\expandafter\let\csname oldend#1\expandafter\endcsname\csname end#1\endcsname
	\renewenvironment{#1}%
	{\linenomath\csname old#1\endcsname}%
	{\csname oldend#1\endcsname\endlinenomath}}%
\newcommand*\patchBothAmsMathEnvironmentsForLineno[1]{%
	\patchAmsMathEnvironmentForLineno{#1}%
	\patchAmsMathEnvironmentForLineno{#1*}}%
\AtBeginDocument{%
	\patchBothAmsMathEnvironmentsForLineno{equation}%
	\patchBothAmsMathEnvironmentsForLineno{align}%
	\patchBothAmsMathEnvironmentsForLineno{flalign}%
	\patchBothAmsMathEnvironmentsForLineno{alignat}%
	\patchBothAmsMathEnvironmentsForLineno{gather}%
	\patchBothAmsMathEnvironmentsForLineno{multline}%
}

\makeatletter

\makeatletter

\begin{document}

\title{A unified framework for imitation dynamics on higher-order networks}

\author{Bingxin Lin$^{1}$, Qingkai Yang$^{1}$, Xianlin Zeng$^{1}$, Jie Chen$^{2,3}$, Hao Fang$^{1, \ast}$, Lei Zhou$^{1, \ast}$
	\\
	\footnotesize{$^{1}$School of Automation, Beijing Institute of Technology, Beijing 100081, China}\\
	\footnotesize{$^{2}$School of Astronautics, Harbin Institute of Technology, Harbin 150001, Heilongjiang Province, China}\\
	\footnotesize{$^{3}$National Key Lab of Autonomous Intelligent Unmanned Systems, School of Automation, Beijing Institute of Technology, Beijing 100081, China}\\
	\footnotesize{$\ast$ Corresponding authors. E-mail: fangh@bit.edu.cn; leizhou@bit.edu.cn}\\
}

\maketitle

\begin{abstract}
Cooperation is central to human societies and often unfolds within groups. 
Higher-order networks, such as hypergraphs, naturally represent these groups as hyperedges. 
Network structures and update rules, by which individuals revise their strategies, are the two fundamental components that shape the evolution of cooperation in structured populations.
Yet while the effects of network structure have been studied extensively, update rules have been examined mostly through isolated models, leaving their relationships and the origins of their differing evolutionary outcomes poorly understood.
Here we develop a unified framework for imitation dynamics on higher-order networks, parameterizing imitation-based update rules by the number of groups an individual samples and the number of peers consulted within each group.
Under weak selection, we derive a closed-form condition for the success of cooperation in any multiplayer social dilemma on homogeneous hypergraphs, encompassing games with both linear and nonlinear payoff structures. 
The framework places previously disconnected update rules within a single family and reduces their effects on cooperation to one interpretable quantity, which we term information diversity. Update rules inducing higher information diversity promote cooperation more effectively, and we prove that this ordering holds strictly across the entire space of multiplayer social dilemmas. 
Simulations extend this principle to heterogeneous hypergraphs constructed both synthetically and from empirical data. Our framework provides a systematic way to represent, analyze, and compare update rules on higher-order networks, turning a fragmented collection of microscopic updating mechanisms into a tractable and interpretable theory.
\end{abstract}
\section*{Introduction}
Cooperation, a prosocial behavior that benefits the collective at a personal cost, is a cornerstone of human societies\cite{nowak1992evolutionary, nowak2006five, ohtsuki2006Nature, allen2017evolutionary}. 
It underpins our capacity to confront global challenges such as mitigating climate change, conserving natural resources, and protecting public health. 
Yet the evolution of cooperation is persistently threatened by the temptation to defect, as individuals can exploit others' cooperation without bearing the associated costs\cite{szabo1998evolutionary,ifti2004effects}. 
Evolutionary game theory provides a powerful framework to study how cooperation emerges and persists in populations of self-interested individuals by explicitly modeling strategic interactions and selection\cite{nowak2004emergence, taylor2007evolution, tarnita2009strategy}. 
Within this framework, cooperation can be favored in network-structured populations, where individuals interact along social ties and update their strategies by imitating successful peers\cite{szabo2007evolutionary, allen2014games, allen2017evolutionary}.
The underlying mechanism that facilitates cooperation is network reciprocity: local interactions allow cooperators to form clusters that shield them from exploitation by defectors and thereby sustain cooperation at the population level\cite{ohtsuki2006Nature,nowak1992evolutionary}.
Importantly, both network structures and update rules, which specify how individuals gather social information and translate it into strategy revisions, shape the evolution of cooperation\cite{nowak2006five,wang2013interdependent,szolnoki2015conformity}.
	
Despite substantial progress on network reciprocity, most studies adopt pairwise interaction structures, which are limited in capturing the complexity of real-world cooperation that unfolds in larger groups\cite{szabo2007evolutionary, allen2014games, ohtsuki2006Nature, allen2017evolutionary,santos2008social,mcavoy2020social,taylor2007evolution}.
For instance, a hunting wolf pack is typically composed of six to ten individuals, and the passage of a United Nations resolution is decided by a vote among all member states (over one hundred and ninety)\cite{zimen1976regulation,mingst2022united,kauffman2007landscape}.
Moreover, empirical research shows that many social interactions naturally involve multiple individuals participating simultaneously, such as discussions of family planning and friendships within university clubs\cite{cencetti2021temporal,goldstone2009collective,goldstone2005computational,chelaru2021high,alvarez2021evolutionary,zachary1977information}.
Higher-order networks, such as hypergraphs, generalize links to hyperedges that can connect multiple individuals\cite{perc2013evolutionary, levine2017beyond,iacopini2024temporal}, making them well-suited to represent group interactions \cite{cencetti2021temporal,benson2018simplicial,battiston2020networks,zhang2023higher,contisciani2022inference}. 
Beyond providing a more faithful representation of real-world interactions, higher-order networks can generate collective dynamics that cannot be reproduced by any combination of pairwise interactions. This irreducibility of higher-order interactions is especially pronounced in the presence of nonlinear effects\cite{battiston2025higher,lotito2024exact}.
Consistent with this perspective, higher-order interaction structures have been shown to qualitatively alter system behavior across a range of collective processes, including oscillator synchronization and epidemic spreading\cite{skardal2021higher,st2021universal,levine2017beyond,gambuzza2021stability,ganmor2011sparse,malizia2025hyperedge}. 
For the evolution of cooperation, recent studies show that conjoined community structure and hyperedge overlaps on higher-order networks contribute positively to cooperation\cite{sheng2024strategy,wang2025emergence}. 

So far, research into cooperation on higher-order networks has paid far more attention to the effects of network structures than to update rules, leaving the role of update rules and their interplay with group structures underexplored. This gap is consequential. On the one hand, since the influence of network structures on evolutionary outcomes depends sensitively on the choice of the update rule\cite{ohtsuki2006Nature,zhou2021aspiration, wang2023imitation}, studying structure in isolation can yield incomplete or even misleading conclusions, and whether a given higher-order structure promotes cooperation may be underspecified until the update rule is fixed. On the other hand, neglecting update rules overlooks a potentially powerful lever, since the way an update rule gathers social information across groups may unlock regimes of enhanced cooperation. Based on these, we take a distinct perspective and ask what general feature of an update rule shapes cooperation on higher-order networks.

To address this, we introduce a class of imitation-based update rules on hypergraphs that exploit group structure, parameterized by the number of hyperedges (groups) an individual samples and the number of peers it consults within each. Under weak selection, we derive a closed-form condition for the success of cooperation in any multiplayer social dilemma on homogeneous hypergraphs, in which every individual belongs to the same number of groups of equal size.
This condition applies to games not only with linear but also with nonlinear payoff structures. In contrast to approaches that fix the update rule and vary structure, we hold the structure fixed and vary a whole family of update rules, and find that the rule's entire effect on cooperation collapses to a single scalar. This scalar captures how an update rule distributes its sampling across groups, a feature we term information diversity and quantify by the probability that two randomly consulted peers come from different groups. 
Specifically, update rules that induce higher information diversity promote cooperation more effectively. 
We analytically and numerically demonstrate this finding in three canonical social dilemmas, namely, the linear public goods game, the multiplayer snowdrift game, and the threshold public goods game. Furthermore, we prove that this ordering holds across the entire space of multiplayer social dilemmas, and extensive simulations show that the principle extends to heterogeneous hypergraphs constructed both synthetically and from empirical data. Together, these results identify information diversity as the feature of update rules that governs cooperation under imitation-based update rules. Our work thus suggests that how individuals gather and combine decision-relevant information across groups is a lever for cooperation, a potential design space for improving social welfare in group-structured systems.

\section*{Results}
\subsection*{Higher-order networks and social dilemmas}
\begin{figure}
	\centering
	\includegraphics[width= \linewidth]{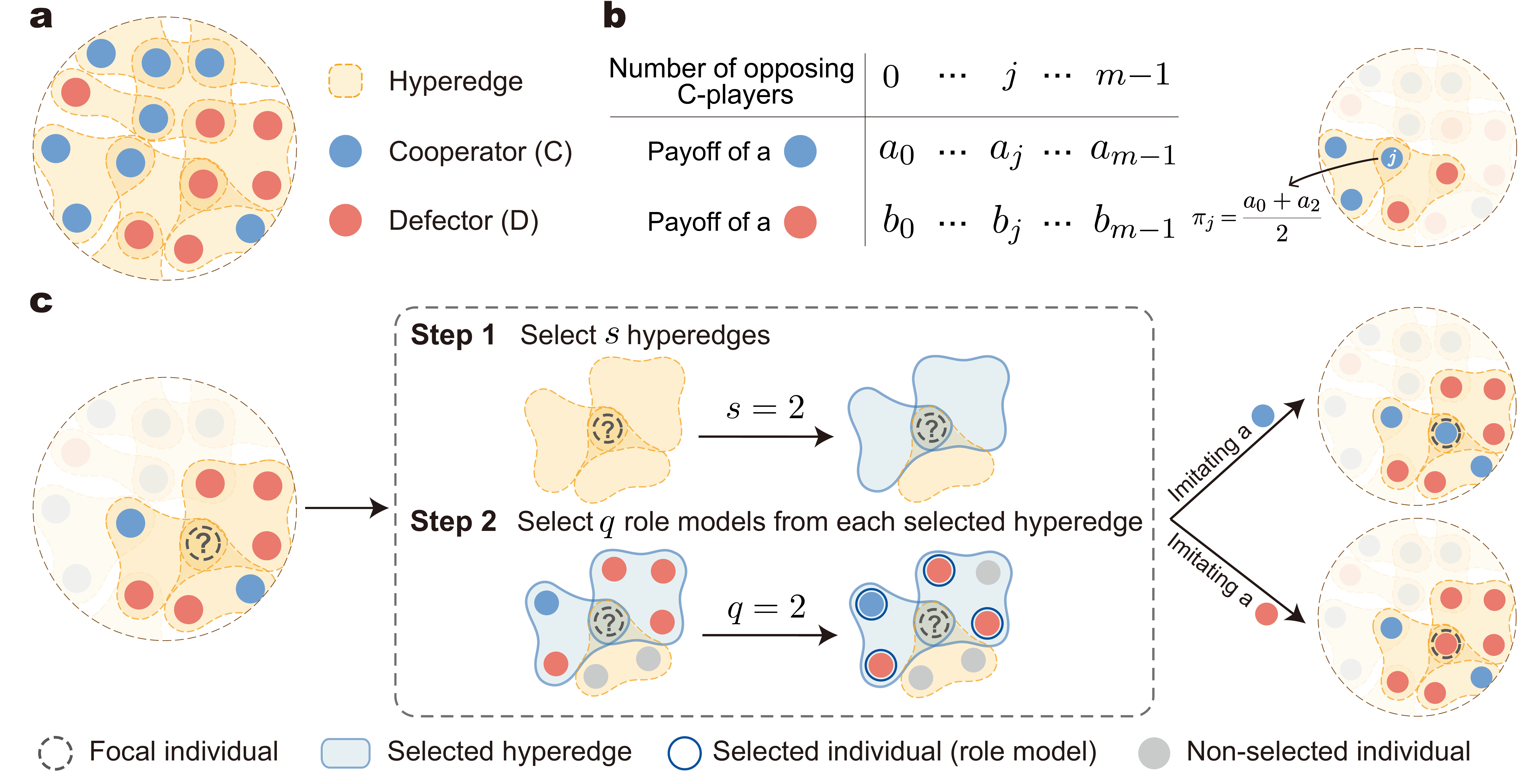}
	\caption{\textbf{Illustration of evolutionary dynamics with parameterized update rules on higher-order networks.}  \textbf{a}, We show a schematic illustration of a higher-order network (hypergraph), which is composed of nodes (marked by solid circles) and hyperedges (marked by light-yellow shaded areas). Here, nodes represent individuals and hyperedges the range of higher-order interactions (i.e., the set of individuals that engage in a higher-order interaction). In addition, each individual can choose either to cooperate or to defect. Correspondingly, they are called cooperators (marked by blue circles) or defectors (marked by red circles).  \textbf{b}, We present a general payoff matrix of an $m$-player game that occurs on hyperedges of size $m$. In such a game, when a cooperator (defector) faces $j$ cooperators among the remaining $m-1$ players, its payoff is denoted as $a_j$ ($b_j$).  On a hypergraph, every individual engages in multiplayer games across all hyperedges it belongs to. The (overall) payoff of an individual, $\pi$, is calculated as the average payoff over all the games it participates in. As illustrated, individual $l$ chooses to cooperate in the two games organized in the hyperedges it belongs to. In these two games, individual $l$ faces $0$ and $2$ other cooperators among its neighbors, respectively. Thus, the payoff of $l$ is $\pi_l=(a_0+a_2)/2$. \textbf{c}, After game interactions, a random individual $l$ is chosen as the focal individual (marked by a dashed circle) to update its strategy. It forgoes its own strategy, selects several role models from its neighbors, and tries to imitate one of them. For the role-model selection process, we explicitly take the higher-order (i.e., group) structure of hypergraphs into account, and it proceeds in two steps: in the first step, the focal individual randomly chooses $s$ hyperedges (marked by light-blue shaded regions) from its $k_l$ hyperedges ($1\le  s \le  k_l$); in the second step, it selects $q$ role models (marked by blue outlined circles) randomly from each chosen hyperedge ($1 \le q \le n-1$), where $n$ denotes the smallest group size among the $k_l$ hyperedges. After all $sq$ role models are selected (non-selected individuals are marked by grey circles), the focal individual then adopts the strategy of one of them with a probability proportional to their fitness. Here, $s$ and $q$ parameterize an imitation process, and each pair $(s,q)$ represents a specific update rule.}\label{fig1}
\end{figure}

	We consider a population of $N$ individuals. These individuals interact through a hypergraph, a representative higher-order network structure consisting of nodes and hyperedges\cite{civilini2021evolutionary}. On hypergraphs, each individual occupies a node, and each hyperedge connects a group of nodes. Unlike traditional pairwise networks where links can only connect a pair of nodes, hyperedges can simultaneously connect an arbitrary number of nodes. The number of nodes in a hyperedge is defined as the group size or the order of the hyperedge. When all hyperedges are of size two, the hypergraph reduces to a pairwise network. In our model, we denote the set of group sizes of all hyperedges as $\mcal{M}=\{m_1,m_2,\ldots ,m_h\}$, where $h$ is the number of hyperedges and $m_j$ is the group size of the $j$-th hyperedge. Analogous to the degree of a node on pairwise networks, we define the number of hyperedges that a node belongs to as its hyperdegree. The set of hyperdegrees of all nodes is denoted as $\mcal{K}=\{k_1, k_2, \ldots, k_N\}$, where $k_j$ is the hyperdegree of node $j$. Fig. \ref{fig1}a shows a schematic of a hypergraph. 

	To explore the evolution of cooperation on higher-order networks, we use multiplayer social dilemmas to capture higher-order interactions among individuals\cite{fu2007social,hilbe2014cooperation}. On hypergraphs, every group of individuals connected by a hyperedge participates in the same social dilemma, and each individual $i$ in total engages in $k_i$ social dilemmas, one for each hyperedge it belongs to. In these social dilemmas, individuals can choose one of two strategies with opposing incentives: cooperation, which promotes collective benefits, and defection, which prioritizes personal gains. Individuals adopting these strategies are called cooperators and defectors, respectively. Here, we adopt a general game model for multiplayer social dilemmas\cite{du2014aspiration,hilbe2014cooperation}, where an individual's payoff in an $m$-player game is determined by how many cooperators are present among the remaining $m-1$ co-players. Specifically, when there are $j$ ($j=0,1,\cdots,m-1$) cooperators in the rest of the group, a cooperator's payoff is $a_j$, and a defector's payoff is $b_j$. Fig. \ref{fig1}b presents the payoff matrix for a general multiplayer game.  In addition, to constitute a social dilemma, the payoffs $a_j$ and $b_j$ ($j=0,1,\cdots,m-1$) must satisfy the following three conditions\cite{kerr2004altruism,pena2016ordering}: 
	(i) individuals always prefer having more cooperative co-players, regardless of their own strategy, i.e., for $0\le j \le m-2$, $a_{j+1} \ge a_j$ and $b_{j+1} \ge b_j$; 
	(ii) in any mixed group, defectors earn no less than cooperators and have strictly higher payoffs in at least one group configuration, i.e., for $0\le j \le m-2$, $b_{j+1} \ge a_j$ and for at least one $j=j^*$, $b_{j+1} > a_{j}$; 
	(iii) mutual cooperation yields a higher payoff than mutual defection, i.e., $a_{m-1} > b_0$. 

	According to the above definition, the number of possible social dilemmas is vast. To illustrate our framework, we consider three canonical multiplayer social dilemmas: the linear public goods game (LPGG)\cite{alvarez2021evolutionary,wang2025emergence}, the multiplayer snowdrift game (MSG)\cite{hauert2004spatial,souza2009evolution}, and the threshold public goods game (TPGG)\cite{wang2009emergence, du2014aspiration}. These games are widely used benchmarks in evolutionary game theory and capture both linear and nonlinear features of cooperation observed in real-world settings\cite{du2014aspiration}. In detail, the LPGG captures a linear accumulation of collective welfare with the number of contributors, whereas the MSG and TPGG capture nonlinearities arising from cost sharing and threshold (critical-mass) effects, respectively. In an LPGG, individuals decide whether to contribute to a public good that increases linearly with the number of contributors and is shared by all group members including non-contributors; in an MSG, individuals decide whether to share the workload of generating a common benefit; and in a TPGG, the public good is generated only if the number of contributors reaches a prescribed threshold, and otherwise all group members lose their endowments.
	
	Consider an $m$-player game where a focal individual faces $j$ cooperators among the remaining $m-1$ players. For the LPGG, the payoffs for the focal individual as a cooperator (C) and a defector (D) are 
	\begin{equation}
		a_j=\frac{(j+1)r_1c}{m}-c,~b_j=\frac{jr_1c}{m},
	\end{equation}
	respectively, where $r_1$ (\(1 < r_1 < m\)) is the multiplication factor in the LPGG. For the MSG, the payoffs are
	\begin{equation}
		a_j=r_2 c-\frac{c}{j+1},~b_j=\left\{
		\begin{array}{ll}r_2 c,& \hbox{$j\neq 0$} \\0, & \hbox{$j=0$}\end{array}\right. ,
		\label{MSG}
	\end{equation}
	where $r_2>1$ is the synergy factor in the MSG. For the TPGG,  the payoffs are 
	\begin{equation}
		a_j=\left\{
		\begin{array}{ll}	\frac{(j+1)r_3c}{m},& \hbox{$j\ge d-1$} \\0, & \hbox{$j<d-1$}\end{array}\right.,~b_j=\left\{
		\begin{array}{ll}\frac{jr_3c}{m}+c,& \hbox{$j\ge d$} \\0, & \hbox{$j< d$}\end{array}\right.,
		\label{TPGG}
	\end{equation}
	where $r_3$ is the multiplication factor and $d$ is the threshold ($0<d<m$) in the TPGG.  Note that all the above games satisfy the conditions required for multiplayer social dilemmas. 
	\subsection*{Update rules and evolutionary dynamics} 
	After game interactions on hypergraphs, each individual $i$ calculates its payoff $\pi_i$ by averaging gains from all games it participates in. 
	Then, it adjusts its strategy according to the update rule specified.
	We assume that individuals use imitation-based update rules, where they update their strategies by imitating successful role models. 
    However, existing imitation-based rules on higher-order networks are typically defined as a handful of disconnected variants, preventing a unified view of how the general feature of update rules shapes evolutionary outcomes.
    To provide a systematic framework, we decompose the imitation process into two elementary steps and propose a class of imitation-based update rules governed by two parameters: the number of hyperedges each individual samples, $s$, and the number of peers consulted in each sampled hyperedge, $q$.

	Under the proposed imitation-based update rules, the evolutionary process proceeds as follows. At each time step, a random individual $l$ is selected as the focal individual.
	It then randomly selects $s$ hyperedges ($1 \le s \le k_l$) from its $k_l$ hyperedges. 
	After that, from each selected hyperedge, it randomly chooses $q$ ($1 \le q \le n-1$) neighbors as role models, where $n$ is the smallest group size among its $k_l$ hyperedges. Here, we assume that all individuals use the same fixed pair of parameters, $s$ and $q$, when updating their strategies.
	We denote the set of role models as $\Omega_l^{(s,q)}$ ($l \notin \Omega_l^{(s,q)}$). 
	The probability that $l$ imitates the strategy of one of its role models $j$ in $\Omega_l^{(s,q)}$ is
	\begin{equation}
		p^{(s,q)}_{l\rightarrow j}=\frac{f_j}{\sum_{i\in \Omega_l^{(s,q)}}f_i},
	\end{equation}
	where $f_j=\exp(w\pi_j)$ is the fitness of individual $j$ and $w$ the intensity of selection\cite{ohtsuki2006Nature,allen2017evolutionary}. 
	Note that when $s=1$ and $q=1$, the focal individual randomly selects a hyperedge, and it imitates a role model randomly chosen in the hyperedge with certainty.  In this case, the evolutionary process essentially corresponds to the neutral drift, where the payoffs have no effect on the evolutionary outcomes. 
	For convenience, we refer to different update rules by their corresponding parameter pairs $(s, q)$ in the following analysis. Fig. \ref{fig1}c illustrates the imitation process governed by the two parameters.
	
	The strategy update process is iterated until the population eventually reaches one of the two absorbing states, full cooperation (the state where everyone cooperates) and full defection (the state where everyone defects). To quantify the influence of different update rules on the evolution of cooperation, we calculate the fixation probabilities of cooperators and defectors. Let $p\in(0,1)$ denote the initial fraction of mutants. When $Np$ cooperators are initially introduced into a population of defectors, the probability that they eventually take over the entire population is the fixation probability of cooperators. Similarly, the fixation probability of defectors is the probability that $Np$ defectors introduced into an all-cooperator population eventually dominate. We denote such fixation probabilities as $\phi_C(p)$ and $\phi_D(p)$, respectively. 
Here, we consider weak selection ($0<w\ll 1$), where the payoff of games has a minor influence on fitness\cite{nowak2004emergence,ohtsuki2006Nature}. To determine whether cooperation is favored over defection by natural selection, we derive a condition for the success of cooperation, namely, an inequality under which $\phi_C(p) > \phi_D(p)$. For the three canonical social dilemmas we consider, such a condition reduces to telling whether the game parameter $r_i$ exceeds its associated critical value $r_i^*$ ($i=1$ corresponds to the LPGG; $i=2$ MSG, and $i=3$ TPGG). A smaller critical value indicates that the model setting is more conducive to the evolution of cooperation. For instance, in the LPGG, if $r_1^*$ under update rule $(s_1, q_1)$ is lower than that under $(s_2, q_2)$, then $(s_1, q_1)$ is more favorable for cooperation. Similarly, $r_2^*$ and $r_3^*$ allow comparisons across update rules in the MSG and TPGG. For brevity, all three critical values are collectively denoted by $r^*$ where appropriate.

\subsection*{Condition for the success of cooperation}
We start by exploring the condition under which cooperation is favored over defection (i.e., $\phi_C(p)>\phi_D(p)$).
Although the condition for the success of cooperation on heterogeneous hypergraphs has been derived, they are implicit and rely on the solution to a system of linear equations of size $O(N^{L+1})$, where $L$ is the largest hyperedge order\cite{sheng2024strategy}. This makes it difficult to isolate the effects of update rules on the evolution of cooperation. To uncover general features of update rules that shape evolutionary outcomes, we therefore focus on homogeneous hypergraphs, where each node belongs to $k$ hyperedges and each hyperedge contains $m$ nodes.
On such a hypergraph $\mcal{H}$, every individual interacts with $m-1$ others within each of their $k$ hyperedges. 

Let $t=(k-1)(m-1)-1$. Under weak selection $0<w\ll 1$, we analytically derive that when $t>0$ and $sq \neq 1$, cooperation is favored over defection if
\begin{equation}\label{phiC-phiD>0}	
	\sum_{j=0}^{m-1}\left[\eta_{(s, q)} F_j(p)+G_j(p)\right]  \left( a_j-b_{m-1-j}\right) >0,
\end{equation}
where 
\begin{equation}\label{eta}
	\eta_{(s, q)} = \frac{k(m-1)}{\left(\frac{sq-1}{sq-q}-1 \right) (k-1)(m-1)+ 1} + 1. 
\end{equation}
Note that when $s = 1$, the above expression is invalid since $sq-q=0$. In this case, we instead set $\eta_{(1, q)}=\lim_{s\to 1} \eta_{(s, q)}=1$. Here, $\eta_{(s,q)}$ is determined by the network structure and the update rule. Importantly, the parameters $s$ and $q$ enter $\eta_{(s, q)}$ only through the single quantity $(sq-1)/(sq-q)$. 
We show in the next subsection that this quantity measures the diversity of information consulted during strategy updating.
In condition (\ref{phiC-phiD>0}), the coefficients $F_j(p)$ and $G_j(p)$ are both polynomial functions of $p$ and symmetric about $p=1/2$ (see equations (\ref{F}) and (\ref{G}) in Methods). 
They are determined by the network structure ($k$ and $m$) and the initial fraction of mutants $p$, but independent of the payoff values $a_j$ and $b_j$ and the update rule. 
We further prove that both $F_j(p)$ and $G_j(p)$ are non-negative for $0\le j \le m-1$ and $p\in (0,1)$ (in particular, $G_{m-1}(p)=0$), and $\eta \ge 1$ whenever $sq \neq 1$ (see Supplementary Section 5.1; hereafter, we abbreviate $\eta_{(s,q)}$ as $\eta$). 

We now provide an intuitive understanding of condition~(\ref{phiC-phiD>0}), which can be equivalently rewritten as
\begin{equation}
\int_0^p \int_v^{1-v}\sum_{j=0}^{m-1}\left[q_{jC|CD}(u)+\eta (k-1)q_{jC|C}(u) \right] (a_j-b_{m-1-j})~dudv>0.
\end{equation}
The condition compares the payoffs of the two strategies in complementary local compositions. A cooperator interacting alongside $j$ other cooperators receives payoff $a_j$, whereas a defector in the complementary composition, with $j$ defecting neighbors, receives payoff $b_{m-1-j}$. The quantity $a_j-b_{m-1-j}$ therefore measures the relative advantage of cooperation across these mirrored environments and is often referred to as the ``gains from flipping''\cite{pena2016ordering}. The coefficient multiplying each payoff difference represents the cumulative contribution of the corresponding local composition over the fixation process. 
To see where this coefficient comes from, consider a cooperator acting as a role model. Its payoff has two parts. The first comes from the hyperedge shared by the role model and the focal defector selected for updating, weighted by $q_{jC|CD}(u)$ (see Methods and Supplementary Section 5.4), the probability that this shared hyperedge contains $j$ further cooperators given the presence of a $CD$ pair, with $q_{(m-1)C|CD}(u)=0$. The second comes from the role model's remaining $k-1$ hyperedges and is weighted by $q_{jC|C}(u)$, the probability that a hyperedge containing the role model also contains $j$ other cooperators.
Both $q_{jC|C}(u)$ and $q_{jC|CD}(u)$ are polynomial functions of $u$. Finally, the double integral indicates that these local compositions are not evaluated at a single population state but are accumulated over the evolutionary trajectory.

A key implication of condition (\ref{phiC-phiD>0}) is that given the network structure, the impact of the update rule on the condition for the success of cooperation is fully encapsulated in $\eta$. 
The significance of this is two-fold: (i) it implies that if different update rules lead to the same value of $\eta$, they have exactly the same impact on the evolution of cooperation; for instance, the update rule with ($s = 4, q = 2$) and ($s = 5, q = 3$) yield the same condition for the success of cooperation; (ii) the classical $\sigma$ rule\cite{tarnita2009strategy,wu2013dynamic} states that, for any $m$-player game, the condition for the success of cooperation is an $m$-parameter inequality $\sum_{j=0}^{m-1}\sigma_j (a_j-b_{m-1-j})>0$, where $\sigma_j$ depends on the population structure and the update rule. However, it does not reveal how these factors jointly determine $\sigma_j$. In contrast, our work supplies an explicit expression, $\sigma_{j}(p)=\eta F_j(p)+G_j(p)$, valid for general multiplayer games, in which the update rule enters only through the single scalar $\eta$, which rescales the structure-relevant coefficient $F_j(p)$ relative to $G_j(p)$.

	Besides, our analytical condition recovers previous results on pairwise networks as special cases. For instance, substituting $m = 2, q=1$ into condition (\ref{phiC-phiD>0}), we can simplify it and obtain that $\phi_C > \phi_D$ if $(k+1)a_1+(k-1)a_0-(k-1)b_1-(k+1)b_0>0$. This condition for strategy success is independent of $s$. Furthermore, in the context of the donation game (i.e., $a_0=-\mcal{C}$, $a_1=\mcal{B}-\mcal{C}$, $b_0=0$, and $b_1=\mcal{B}$), this condition reduces to the well-known rule $\mcal{B}/\mcal{C}>k$\cite{ohtsuki2006Nature}.

	To gain more intuition about the effect of update rules on the evolution of cooperation, we apply condition (\ref{phiC-phiD>0}) to the three social dilemmas we mentioned earlier, namely, LPGG, MSG, and TPGG.
	For the LPGG, we derive that cooperation is favored over defection whenever $r_1> r_1^*$, where the critical value
	\begin{equation}\label{r1*}
		r_{1}^*=\frac{m}{k\eta}+\frac{m(k-1)}{k}.
	\end{equation}
	Similarly, for the MSG and TPGG, their corresponding critical value $r_{2}^*$ and $r_{3}^*$ are
	\begin{equation}\label{r2*}
		r_{2}^*=\frac{ \sum_{j=0}^{m-1}\frac{\eta F_j(p)+G_{j}(p)}{j+1}}{\eta F_{m-1}(p)},
	\end{equation}
	and
	\begin{equation}\label{r3*}
		r^*_{3}=\frac{m \sum_{j=d-1}^{m-2}\left(\frac{G_{j-d+1}(p)}{\eta}+F_{j-d+1}(p) \right)}{mF_{m-1}(p) +\sum_{j=d-1}^{m-2} \left[F_j(p)-F_{m-2-j}(p)\right] (j+1)},
	\end{equation}
	respectively. 
\subsection*{Interpretation through information diversity}
Despite the intricate dependence on $F_j(p)$ and $G_j(p)$, we mathematically prove that the critical values $r_1^*$, $r_2^*$, and $r_3^*$ are all decreasing functions of $\eta$, independent of $k,~m,~p$ and $d$ (see Supplementary Section 5.1). This means that when $\eta$ increases, these critical values decrease, resulting in a more favorable condition for the evolution of cooperation in the associated social dilemmas. Moreover, $\eta$ is an increasing function of $(sq-q)/(sq-1)$ (see equation (\ref{eta})), which leads to the fact that as $(sq-q)/(sq-1)$ increases, the critical value for LPGG, MSG, or TPGG falls.
Note that $(sq-q)/(sq-1)$ is only affected by $s$ and $q$ that govern the update rule.
Thus, update rules with higher $(sq-q)/(sq-1)$ promote cooperation more effectively in the LPGG, MSG, and TPGG.

\begin{figure}
	\centering
	\includegraphics[width=\linewidth]{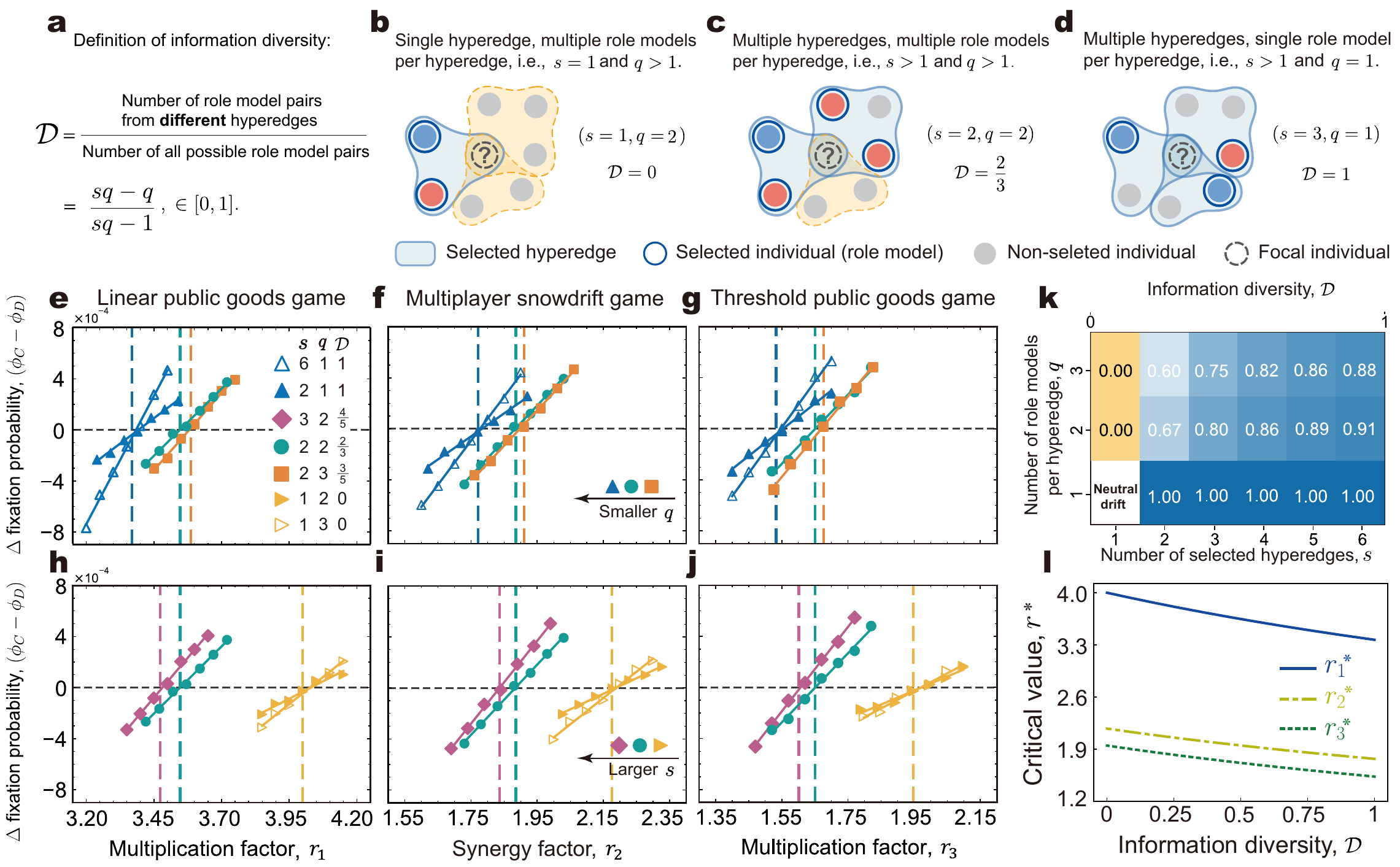}
	\caption{\textbf{Update rules with high information diversity promote the evolution of cooperation.} \textbf{a}, Information diversity $\mcal{D}$ is defined as the probability that any two randomly selected role models among the $sq$ ones come from different hyperedges. \textbf{b-d}, Depending on the values of $s$ and $q$, four distinct cases can be identified: (1) $s=1$, $q>1$;  (2) $s>1$, $q>1$; (3) $s>1$, $q=1$; and (4) $s=1$, $q=1$. The first three cases correspond to panels \textbf{b-d}, which respectively present the values of information diversity $\mcal{D}$ under each setting. The fourth case is excluded in our study since it represents neutral drift. \textbf{e-j}, We plot the difference between the fixation probability of cooperators $\phi_C$ and that of defectors $\phi_D$, i.e., $\phi_C-\phi_D$, as a function of the key game parameter in three distinct social dilemmas, which are linear public good games (LPGG, \textbf{e,h}),  multiplayer snowdrift games (MSG, \textbf{f,i}), and threshold public goods games (TPGG, \textbf{g,j}). Under these games, we examine update rules with seven $(s,q)$ combinations: $(6,1),~(2,1),~(3,2),~(2,2),~(2,3),~(1,2)$, and $(1,3)$. In detail, panels \textbf{e}, \textbf{f}, and \textbf{g} show the effect of the number of role models from each hyperedge, $q$ on the evolutionary outcomes. And panel \textbf{h}, \textbf{i}, and \textbf{j} present the effect of the number of selected hyperedges, $s$, on the evolution of cooperation. In these panels, each data point is the fixation probability difference obtained from $10^7$ independent simulations; each solid line is the result obtained by the linear fit to the corresponding data and is used to guide the eye. The vertical dashed lines indicate the critical values predicted by our theoretical results. \textbf{k}, We plot the values of information diversity $\mcal{D}$ for different $(s, q)$ pairs. \textbf{l}, Theoretical analysis reveals that, across all three games, the critical value decreases with increasing information diversity. Parameter values: $N=500$, $k=6$, $m=4$, $d=2$, $p=1/N$, $w=0.01$.\label{fig2}}
\end{figure}

	Based on the crucial role $(sq-q)/(sq-1)$ plays in shaping the evolutionary outcomes, it is natural to ask what it stands for in the context of evolutionary dynamics on higher-order networks. 
	To answer this, consider the information acquisition process during strategy updating: 
	when a focal individual is picked, it first randomly selects $s$ hyperedges it belongs to as the sources of information, then it randomly chooses $q$ individuals from each source to collect information (i.e., strategies and payoffs), and after that, it imitates the strategy of one of the role models with a probability proportional to fitness.
	In this process, $s$ specifies the number of information sources individuals need to visit, and $q$ determines the amount of information acquired from each source.
	We now introduce a new quantity, termed information diversity, and denoted by $\mcal{D}$, as the probability that, among all $s q$ pieces of information, two randomly selected ones originate from different information sources (Fig. \ref{fig2}a). Based on our analysis, the information diversity $\mcal{D}$ associated with the update rule $(s,q)$ is
	\begin{equation}\label{diversity}
		\mcal{D}=\dfrac{\binom{s}{2}\binom{q}{1}\binom{q}{1}}{\binom{sq}{2}}=\frac{sq-q}{sq-1}.
	\end{equation} 
	This means that $(sq-q)/(sq-1)$ can be interpreted as the diversity of consulted information during strategy updating.  Therefore, our results show that increasing the information diversity $\mcal{D}$ during strategy updating lowers the critical value for cooperation to prevail over defection. 
	
	To further investigate how $s$ and $q$ affect the information diversity $ \mcal{D}$, we calculate its partial derivative with respect to $s$ and $q$ ($sq \neq 1$), and get
	\begin{equation}\label{derivative_s}
		\frac{\partial  \mcal{D}}{\partial s}  =  \frac{q(q-1)}{(sq-1)^2}, ~~~\frac{\partial  \mcal{D}}{\partial q} = \frac{1-s}{(sq-1)^2}.
	\end{equation}	
	Based on the signs of these partial derivatives, there are three typical cases shown in Fig.~\ref{fig2}b-d. The first one occurs when $s=1$ and $q>1$. In this case, $\partial \mcal{D}/\partial q=0$ and $\mcal{D}$ reaches its minimum value of $0$, irrespective of $q$, meaning that any pair of role models comes from the same hyperedge. The resulting class of update rules is the least favorable for the evolution of cooperation. The second one occurs when individuals update strategies using information from multiple sources ($s>1$) and collecting multiple pieces of information within each source ($q>1$). For this class, increasing $s$ or reducing $q$ enhances cooperation. For instance, the update rule with $s=3$ and $q=2$ promotes cooperation more effectively than that with $s=2$ and $q=2$. The third one occurs when $s>1$ and $q=1$. Here, $\mcal{D}$ reaches its maximum value of $1$, meaning that any pair of role models comes from different hyperedges.  The resulting class of update rules is the most favorable for the evolution of cooperation. 
	
	To verify our theoretical predictions for critical values in equations (\ref{r1*}), (\ref{r2*}), and (\ref{r3*}), we run a series of simulations and plot the results in Figs. \ref{fig2}e-j under seven different update rules that are parameterized by $(s,q) = (6,1), (2,1), (3,2), (2,2), (2,3), (1,2), (1,3)$. The corresponding values of information diversity are $\mcal{D}=1, 1, 4/5, 2/3, 3/5, 0, 0$, respectively. Here, we use symbols to represent the data obtained via simulations and solid lines for the linear fit to the corresponding data. Moreover, for better illustrations, we represent the theoretical critical value by the vertical dashed line. To separately analyze the effects of $s$ and $q$ on evolutionary outcomes, Figs. \ref{fig2}e-j are organized into two rows. In the first row, we find that decreasing $q$ increases information diversity and leads to a smaller critical value, with the minimum critical value attained at $q=1$. In the second row, decreasing $s$ reduces information diversity and results in a larger critical value, with the maximum critical value occurring at $s = 1$. These conclusions hold consistently across all three games.
	In addition, Fig. \ref{fig2}k illustrates the values of information diversity for the different $(s,q)$ pairs. Note that different $(s, q)$ pairs may yield the same value of $\mcal{D}$; for example, $\mcal{D}=6/7$ for both $(s = 4, q = 2)$ and $(s = 5, q = 3)$. Fig. \ref{fig2}l shows the relationship between information diversity and the theoretical critical value $r_1^*$ in the LPGG, $r_2^*$ in the MSG, and $r_3^*$ in the TPGG, demonstrating that they are all decreasing functions of the information diversity $\mcal{D}$. Here, we set the selection intensity $w=0.01$ in Figs.~\ref{fig2}e-j. To test the robustness of our findings to selection intensities, we run additional simulations for $w=0.05$ and $w=0.1$ (see Supplementary Fig.~1). The results show that our findings remain valid under stronger selection intensities. 
	
	In addition to the baseline setting where individuals acquire $sq$ pieces of social information from their neighbors for imitation, we also consider, in Supplementary Section 4, a variant of the model that incorporates personal information. Specifically, the focal individual $l$ follows the same procedure to sample $sq$ neighbors and additionally considers its own information upon strategy updating. From the perspective of imitation, it considers itself as one of the selected ``role models". 
	In this way, the role models together form a new set $\hat\Omega_l^{(s,q)}$, with $l \in \hat\Omega_l^{(s,q)}$ and $|\hat\Omega_l^{(s,q)}|=sq+1$. Under this setting, we also derive the condition for the success of cooperation. The only difference from condition (\ref{phiC-phiD>0}) is that the coefficient $\eta$ is replaced by $\hat\eta=\frac{k(m-1)}{\left( 1/\mcal{\hat D}-1 \right) (k-1)(m-1)+ 1} + 1$, where
	\begin{equation}
		\mcal{\hat D} =\frac{\binom{s}{2}\binom{q}{1}\binom{q}{1}}{\binom{sq+1}{2}}= \frac{sq-q}{sq+1}.
	\end{equation}
	Notably, the two models differ solely in the expression for information diversity.
	For fixed $s$ and $q$, we always have $\mcal{\hat D}\le \mcal{D}$, implying that incorporating personal information is detrimental to the promotion of cooperation. 
The intuition is that the focal individual belongs to all of its sampled hyperedges, so no focal-neighbor pair is counted as cross-source; the favorable pairs in the numerator are unchanged while the total number of pairs grows, lowering diversity.
With personal information, our model recovers the previous ``imitation" (IM) rule\cite{ohtsuki2006Nature,wang2023imitation,nowak2010evolutionary,allen2014games,allen2017evolutionary} for $s = k$ and $q = m-1$ and the classical ``pairwise comparison" (PC)\cite{wang2023imitation,zhou2021aspiration,szabo1998evolutionary,traulsen2006stochastic} rule for $s = 1$ and $q = 1$. To verify our theoretical predictions, we also conduct simulations under different update rules with personal information. The results in Supplementary Fig.~2 show that the theoretically predicted critical values agree closely with those obtained from simulations. When personal information is considered, increasing either $s$ or $q$ never worsens the outcome, and the conclusion that update rules with higher information diversity lead to lower critical values remains unchanged.

Taken together, our unified framework explains why classical update rules differ in their ability to promote cooperation and identifies information diversity as the feature determining their impact on cooperation.
Since incorporating personal information affects only the value of information diversity, we focus, in the remainder of this work, on evolutionary dynamics without personal information.

\subsection*{Effects of higher-order network structures}
So far, we have focused on homogeneous higher-order networks with $k=6$ and $m=4$. Next, we systematically examine the impact of update rules on the evolutionary outcomes under different higher-order network structures, including homogeneous hypergraphs with other $k$ or $m$, and heterogeneous hypergraphs with non-uniform hyperdegree or order distributions. 
	
\begin{figure}
	\centering
	\includegraphics[width=1\linewidth]{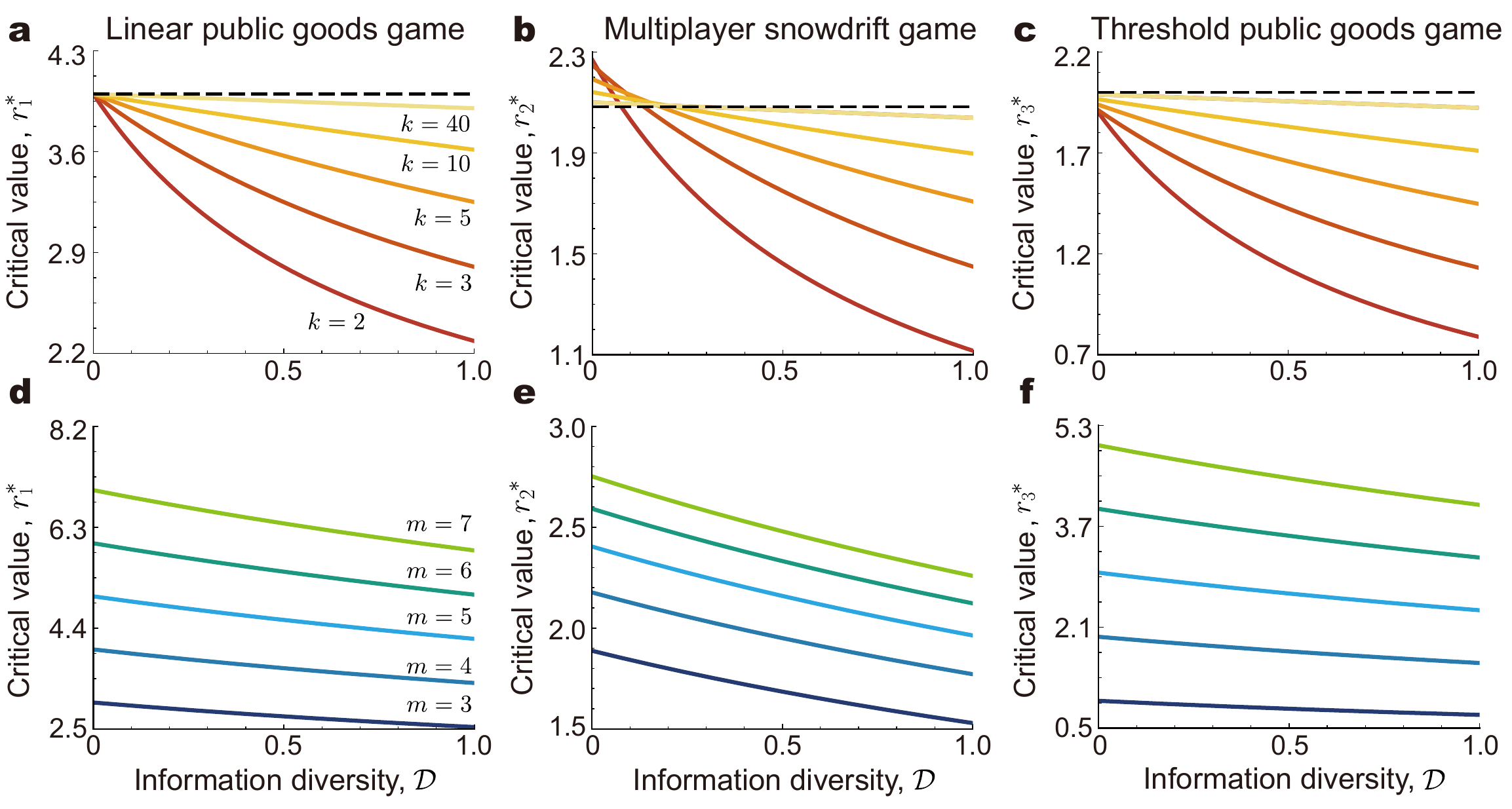}
	\caption{\textbf{Effect of hyperdegree and order (i.e., hyperedge size) on evolutionary outcomes.} The top and bottom panels show the effects of the node hyperdegree $k$ and the hyperedge order $m$, respectively, on the theoretically predicted critical values. Results show that update rules with higher information diversity $\mcal{D}$ consistently promote cooperation across hypergraphs with varying hyperdegree $k$ and order $m$. Here, the black horizontal dashed lines in \textbf{a-c} indicate the theoretical critical values for the $4$th-order complete hypergraph \cite{lin2025evolutionary} in three social dilemmas. Our results show that as the hyperdegree $k$ increases, the hypergraph $\mcal{H}$ approaches a complete hypergraph, and the differences in critical values under different update rules become smaller. Parameter values: $m=4$ (\textbf{a,b,c}), $k=6$ (\textbf{d,e,f}), $d=2$. \label{fig3}}
\end{figure}

	Fig. \ref{fig3} shows how the critical values change as the corresponding information diversity varies under different hyperdegree $k$ and order $m$. 
Overall, for different values of $k$ and $m$, increasing information diversity $\mcal{D}$ monotonically reduces the critical values in LPGG, MSG, and TPGG. 
These findings are consistent with our theoretical results, which demonstrate that the critical value in these social dilemmas decreases with the information diversity $\mcal{D}$, regardless of $k$ and $m$.
	In addition, when examining the effect of hyperdegree $k$ on the critical value (Fig. \ref{fig3}a,b,c), we compare the results for different hyperdegrees with the theoretical critical values for the $4$th-order complete hypergraph $\mcal{H}^{c}_{4\text{th-order}}$ (black horizontal dashed lines), as reported in a recent study\cite{lin2025evolutionary}. Here, $\mcal{H}^{c}_{m\text{th-order}}$ denotes an $m$th-order complete hypergraph in which every set of $m$ nodes forms a hyperedge. 
It serves as the unstructured baseline, namely, the higher-order counterpart of a well-mixed population. This implies that a higher-order network structure is considered to promote cooperation only if its critical value lies below that of the corresponding complete hypergraph.
Notably, in the results shown for the MSG (Fig. \ref{fig3}b), when the information diversity is relatively low, the critical values of higher-order networks with different hyperdegrees $k$ all exceed those of the corresponding complete hypergraphs, indicating that these network structures fail to promote cooperation in the MSG. However, as the information diversity $\mcal{D}$ increases, the critical values can drop below the black dashed line.
This demonstrates that adopting more effective update rules can reverse the effect of network structures on cooperation. It highlights the interplay between network structure and update rules, emphasizing that considering network structure alone is insufficient to fully understand evolutionary dynamics on higher-order networks. 

\begin{figure}
	\centering
	\includegraphics[width=1\linewidth]{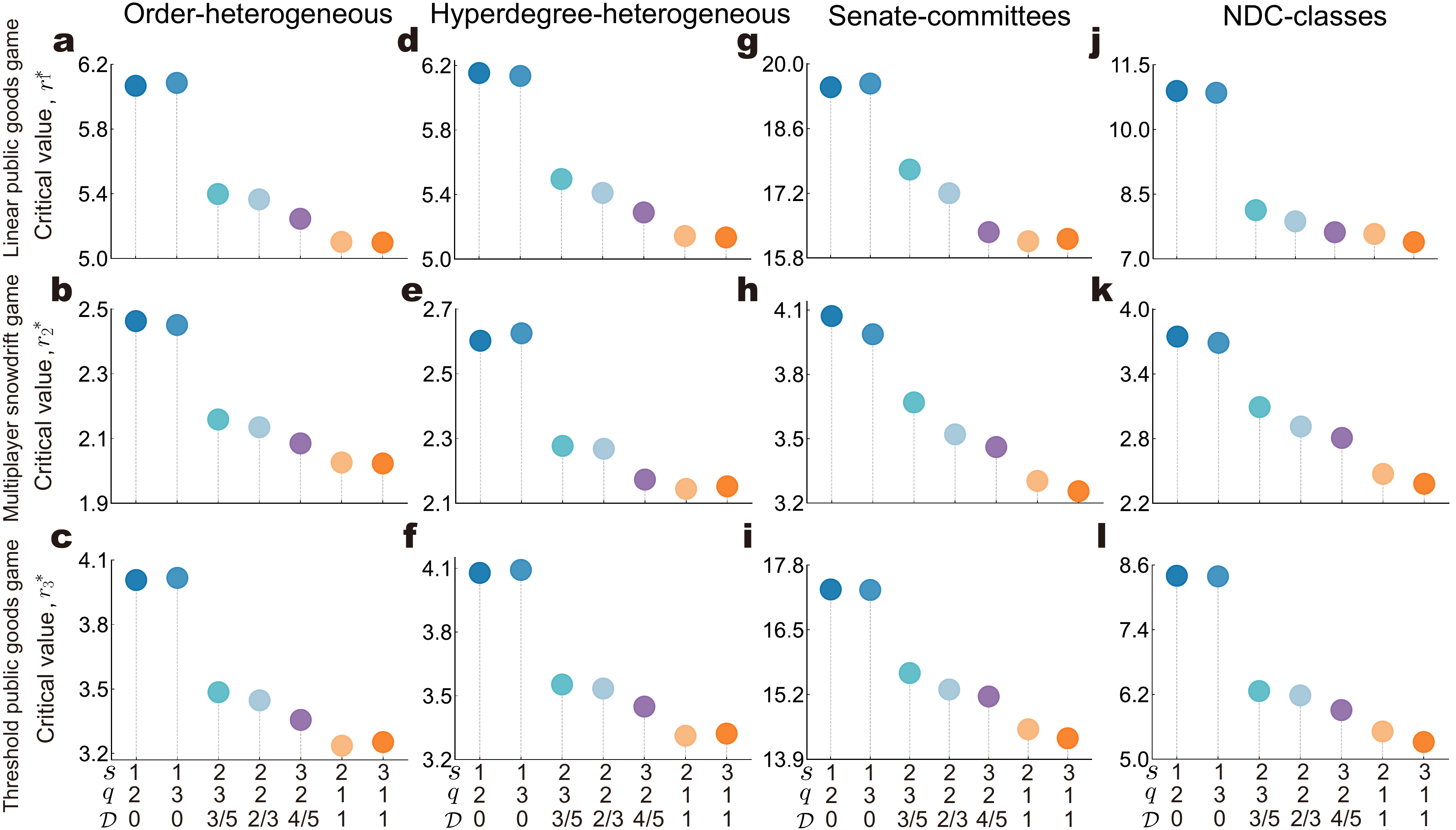}
	\caption{\textbf{Critical values for cooperation on heterogeneous higher-order networks constructed synthetically and from empirical data.} We run a series of simulations and calculate the critical values under various update rules associated with different information diversity $\mcal{D}$ on four heterogeneous hypergraphs: two constructed synthetically and two from real-world datasets. The first-column panels (\textbf{a-c}) show results for synthetic order-heterogeneous hypergraphs, where each node belongs to $k = 6$ hyperedges and the order of hyperedges follows a power-law distribution with a mean of six. The second-column panels (\textbf{d-f}) show results for synthetic hyperdegree-heterogeneous hypergraphs, where all the orders of hyperedges are set to be six while the hyperdegrees follow a power-law distribution with a mean of six. The third-column panels (\textbf{g-i}) show results for the empirical senate committees\cite{landry2024senatecommittees} network with $N=$257 nodes, whereas the fourth-column panels (\textbf{j-l}) show results for an empirical pharmaceutical classification network with $N=$237 nodes from the National Drug Code Directory\cite{landry2023ndcclasses,benson2018simplicial}. Here, to consider update rules with different values of information diversity, we use seven $(s,q)$ combinations: $(1,2),~(1,3),~(2,3),~(2,2),~(3,2),~(2,1)$, and $(3,1)$. Our results indicate that enhancing information diversity during the strategy updating process promotes cooperation on both synthetic and empirical heterogeneous higher-order networks. This means that our findings on homogeneous higher-order networks also apply to non-homogeneous ones. Parameter values: $N=500$ (\textbf{a-f}), $p=1/N$,  $w= 0.01$, $d=2$.\label{fig4}}
\end{figure}

	Besides, we run extensive simulations and investigate the evolution of cooperation on both synthetic and empirical heterogeneous higher-order networks. We plot the critical values as a function of the information diversity $\mcal{D}$ in Fig. \ref{fig4}. Here, we consider four heterogeneous hypergraphs: two types of synthetic heterogeneous hypergraphs and two empirical heterogeneous hypergraphs derived from real-world datasets. The first one is a synthetic order-heterogeneous hypergraph, in which each node has a fixed hyperdegree of $k=6$, while the hyperedge orders follow a power-law distribution with a mean of six (Fig. \ref{fig4}a,b,c). The second one is a synthetic hyperdegree-heterogeneous hypergraph, where each hyperedge has a fixed order of $m=6$, while the hyperdegrees follow a power-law distribution with a mean of six (Fig. \ref{fig4}d,e,f). The choice of power-law distributions for hyperdegree and hyperedge order is motivated by previous empirical findings that show the size and number of groups individuals participate in often follow power-law distributions\cite{milojevic2014principles,newman2001scientific,perra2012activity}. 
The third hypergraph is an empirical senate committees network with $N = 257$ nodes (Fig.~\ref{fig4}g,h,i)\cite{landry2024senatecommittees}. The fourth hypergraph is an empirical pharmaceutical classification network with $N =237$ nodes, derived from the pharmaceutical classes in the National Drug Code Directory (Fig.~\ref{fig4}j,k,l)\cite{landry2023ndcclasses,benson2018simplicial}. Schematic diagrams of these two empirical hypergraphs are provided in Supplementary Fig.~3. In Fig.~\ref{fig4}, the critical values are obtained from simulations under seven update rules with different $s$ and $q$, i.e., $(s,q)=(1,2)$, $(1,3)$, $(2,3)$, $(2,2)$, $(3,2)$, $(2,1)$, and $(3,1)$, which cover five levels of information diversity, namely,  $\mcal{D}=0$, $3/5$, $2/3$, $4/5$, and $1$. Similarly, when information diversity is low, increasing the number of information sources (enlarging $s$) can substantially reduce the critical value for the emergence of cooperation. For example, increasing $s$ from $1$ to $2$ at $q = 2$ leads to the rise of $\mcal{D}$ from $0$ to $2/3$, which reduces the critical value by $11.6\%$, $13.3\%$, and $14.0\%$ for the LPGG, MSG, and TPGG, respectively (Fig.~\ref{fig4}a,b,c). Our results show that enhancing information diversity during strategy updating fosters cooperation not only on homogeneous higher-order networks but also on higher-order networks with heterogeneous structures.    

\subsection*{Extension to general social dilemmas}
	So far, we have explored how information diversity $\mcal{D}$ affects evolutionary outcomes under three specific social dilemmas (LPGG, MSG, and TPGG). Note that, once the network structure and the threshold $d$ (for the TPGG) are fixed, each of these social dilemmas is characterized by a single free game parameter. In these cases, the critical value (e.g., $r^*_1$ in the LPGG) can be solved explicitly and compared between different update rules. A smaller critical value indicates that the corresponding update rule can sustain cooperation across a broader range of social dilemmas. Accordingly, the game space supporting cooperation is essentially one-dimensional, namely, the parameter values above the critical value (e.g., $r_1 > r_1^*$ in the LPGG).
	Given that our condition for the success of cooperation applies to arbitrary social dilemmas on homogeneous higher-order networks (see condition (\ref{phiC-phiD>0})), it is natural to ask whether the effectiveness of update rules in promoting cooperation can still be ordered by the associated values of information diversity in a more general space of social dilemmas. 

	Before we address this, it is important to clarify what it means for one update rule to be more effective in promoting cooperation than the other in the context of general social dilemmas. We denote $\mcal{S}_{\mcal{D}}$ as the set of games where the update rule with information diversity $\mcal{D}$ favors cooperation. In other words, given a homogeneous higher-order network, the update rule with information diversity $\mcal{D}$ always leads to the outcomes that cooperation prevails over defection under any game that belongs to $\mcal{S}_{\mcal{D}}$. 
Based on this, on the same higher-order network, if two update rules with information diversity $\mcal{D}_1$ and $\mcal{D}_2$ result in the relation $\mcal{S}_{\mcal{D}_2} \subsetneq \mcal{S}_{\mcal{D}_1}$, we say that the update rule with information diversity $\mcal{D}_1$ is more effective in promoting cooperation than that with $\mcal{D}_2$. 

We now turn to the condition for the success of cooperation for general social dilemmas. For convenience, we normalize all the game parameters into the interval $[0,1]$ by applying an affine transformation to the general payoff matrix with $a_j$ and $b_j$. The normalized payoff values are $\tilde{a}_j = \frac{a_j - e_2}{e_1 - e_2}$ and $\tilde{b}_j = \frac{b_j - e_2}{e_1 - e_2}$, where $e_1 = \max \bigl\{ a_j, b_j \,\big|\, 0 \le j \le m-1 \bigr\}$ and $e_2 = \min \bigl\{ a_j, b_j \,\big|\, 0 \le j \le m-1 \bigr\}$.
	Note that such a transformation does not change the evolutionary outcome about which strategy eventually prevails. In addition, since the condition for the success of cooperation depends on the difference between payoff values, we can reduce the number of variables to $m$ by defining
	 $\Delta_j=\tilde{a}_j-\tilde{b}_{m-1-j}$.
	In this way, we rewrite the condition for the success of cooperation as
	\begin{equation}\label{simplified-Condition}
	 \sum_{j=0}^{m-1} \sigma_{(s,q,j)}(p)  \Delta_j >0,
	\end{equation}
where $\sigma_{(s,q,j)}(p)=\eta_{(s, q)} F_j(p)+G_j(p)$ and $\mcal{S}_{\mcal{D}}$ is the set of $\Delta_j$ ($j=0,\ldots, m-1$) that satisfies the above condition. Furthermore, to exclude games that are not social dilemmas, we impose the constraints $\Delta_0\le0,~\Delta_0\le\Delta_1\le\Delta_2\le\cdots\le \Delta_{m-1}$, and $\Delta_{m-1}>0$, derived from the first and third condition required by a social dilemma. 
	
\begin{figure}
	\centering
	\includegraphics[width=\linewidth]{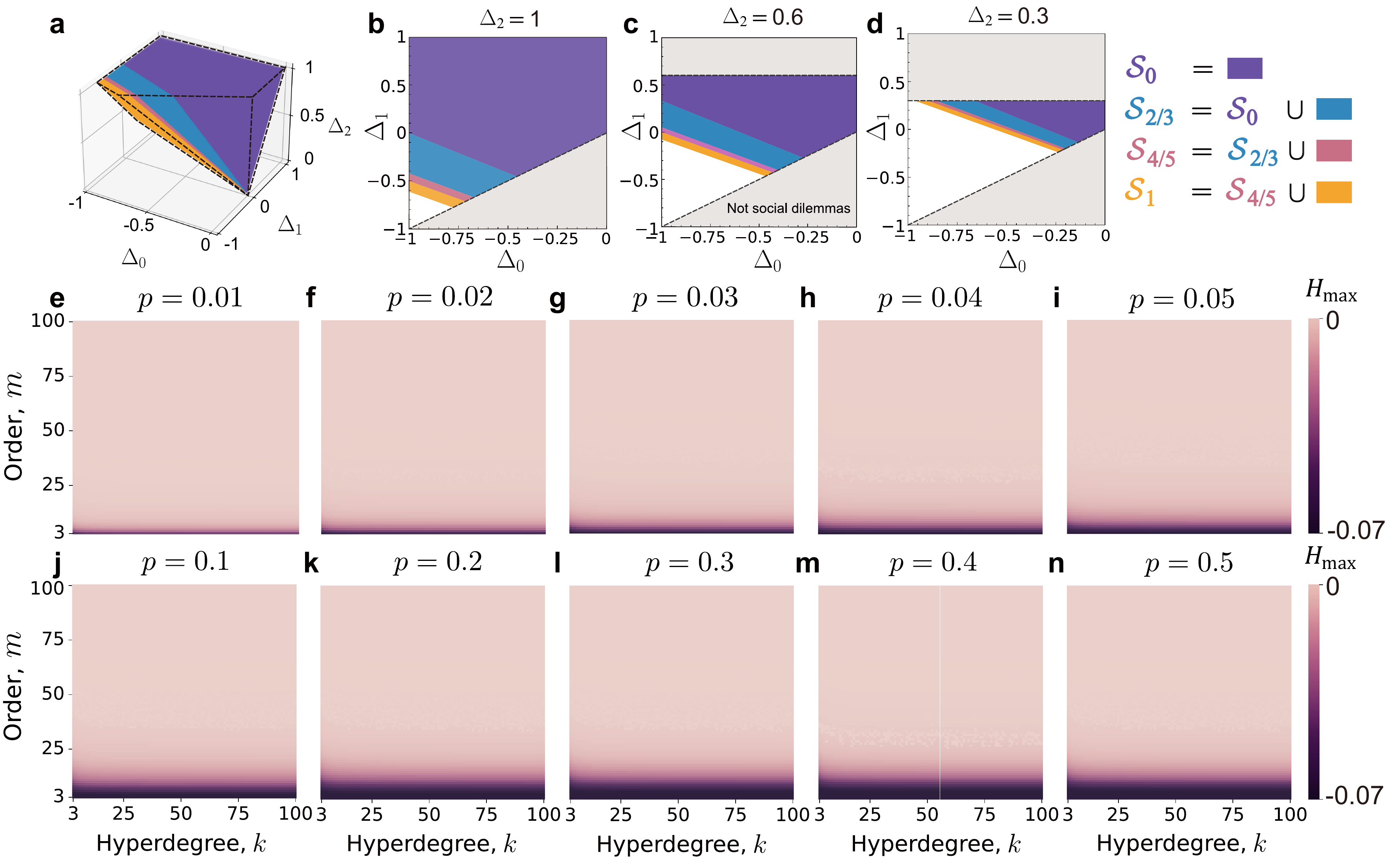}
	\caption{\textbf{Positive impacts of update rules with high information diversity on cooperation in general multiplayer social dilemmas.} We define $\mcal{S}_{\mcal{D}}$ as the set of games (or game space) where cooperation prevails over defection under an update rule with information diversity $\mcal{D}$. To determine whether the update rule with $\mcal{D}_1$ is more effective in promoting cooperation than that with $\mcal{D}_2$ in general social dilemmas, we compare $\mcal{S}_{\mcal{D}_1}$ with $\mcal{S}_{\mcal{D}_2}$: if $\mcal{S}_{\mcal{D}_1} \supsetneq \mcal{S}_{\mcal{D}_2}$, the former one is more effective; otherwise, the latter one is more effective. \textbf{a}, To provide an illustration of $\mcal{S}_{\mcal{D}}$, we take three-player games as an example (i.e., $m=3$) and plot the game space $\mcal{S}_{\mcal{D}}$ that supports cooperation with the information diversity $\mcal{D}=0,~2/3,~4/5,$ and $1$. Here, $\mcal{S}_{\mcal{D}}$ is obtained via linear programming by jointly considering the constraints of the social dilemmas and the condition for the success of cooperation shown in equation (\ref{phiC-phiD>0}). Results show that $\mcal{S}_1\supsetneq \mcal{S}_{4/5}\supsetneq \mcal{S}_{2/3}\supsetneq \mcal{S}_0$. This aligns with our findings that the greater the information diversity, the easier it is for the evolution of cooperation. \textbf{b-d}, The three panels show cross-sections of panel \textbf{a} at $\Delta_2=1,0.6,0.3$, respectively. The grey areas represent games that are not social dilemmas. \textbf{e-n}, If $H_{\mathrm{max}}(k,m,p)< 0$, update rules with higher information diversity promote cooperation more effectively across general $m$-player social dilemmas on hypergraphs characterized by hyperdegree $k$, order $m$, and initial mutant fraction $p$.
		We mathematically prove that $H_{\mathrm{max}}(k,m,p)< 0$ holds for arbitrary $k$, $m$ satisfying $(k-1)(m-1)>1$, and $p\in(0,1)$, with extensive numerical verification across a broad parameter space.
		We evaluate $H_{\mathrm{max}}(k,m,p)$ for $k,m\in[3,100]$ and initial mutant fraction $p\in \{0.01,0.02,0.03,0.04,0.05,0.1,0.2,0.3,0.4,0.5\}$, and find that $H_{\mathrm{max}}(k,m,p)< 0$ across all parameter combinations considered. This means that beyond canonical social dilemmas such as LPGG, MSG, and TPGG, our finding that enhancing information diversity promotes cooperation also holds for general multiplayer social dilemmas. Parameter values: $k=3$, $m=3$, $p=0.1$ (\textbf{a-d}).\label{fig5}}
\end{figure}

	In Fig. \ref{fig5}a, we provide an intuitive illustration of $\mcal{S}_{\mcal{D}}$ on hypergraphs with $k=3$ and $m=3$. As shown, $\mcal{S}_1\supsetneq \mcal{S}_{4/5}\supsetneq \mcal{S}_{2/3}\supsetneq \mcal{S}_0$, which is consistent with the descending order of the associated information diversity: $1>4/5>2/3>0$. Fig. \ref{fig5}b–d show cross-sections of Fig. \ref{fig5}a at $\Delta_2=1,0.6,0.3$, respectively. The grey areas represent the game space that does not satisfy the constraints imposed by being a social dilemma. The white areas indicate the set of games that are social dilemmas but do not support cooperation for the update rules we consider. For a three-dimensional space ($m=3$), we can directly solve for the game space that supports cooperation under different update rules using linear programming and visually compare them. However, this approach becomes infeasible when $m$ is large.
	
For multiplayer social dilemmas with more than three players (i.e., $m>3$), we adopt a different approach\cite{pena2016ordering} that originates from the studies on stochastic orders to determine whether the containment relation $\mathcal{S}_{\mathcal{D}_2} \subsetneq \mathcal{S}_{\mathcal{D}_1}$ holds when $\mathcal{D}_2 < \mathcal{D}_1$. Note that the left-hand side of inequality (\ref{simplified-Condition}) is linear in $\Delta_j$. This implies that the containment relation between the feasible regions $\mcal{S}_{\mcal{D}_1}$ and $\mcal{S}_{\mcal{D}_2}$ is fully determined by the coefficient vectors $\boldsymbol{\sigma}_{(s_1,q_1)}(p)=(\sigma_{(s_1,q_1,1)},\ldots, \sigma_{(s_1,q_1,m-1)})$ and $\boldsymbol{\sigma}_{(s_2,q_2)}(p)$. Based on this, we obtain that the containment relation $\mathcal{S}_{\mathcal{D}_2} \subsetneq \mathcal{S}_{\mathcal{D}_1}$ holds if
	\begin{equation}
	\sum_{j=0}^{x}\hat{\sigma}_{(s_1,q_1,j)}(p) < \sum_{j=0}^{x}\hat{\sigma}_{(s_2,q_2,j)}(p)~\text{for all} ~x=0,1,\cdots ,m-2,
	\end{equation}
	where $	\hat{\sigma}_{(s,q,j)}(p)=\sigma_{(s,q,j)}(p)  / \sum_{i=0}^{m-1} \sigma_{(s,q,i)}(p) $ is the normalized form of $\sigma_{(s,q,j)}(p)$.
	In our model, this inequality can be further reduced (see Supplementary Section 5.2 for details). Specifically, $\mcal{D}_1>\mcal{D}_2$ guarantees $\mcal{S}_{\mcal{D}_1}\supsetneq \mcal{S}_{\mcal{D}_2} $ if 
	\begin{equation}\label{Hx}
		H(k,m,p,x)=\sum_{j=0}^{x} \left[F_j(p)-(k-1)G_j(p)\right]<0~\text{for all} ~x=0,1,\cdots ,m-2,
	\end{equation}
where $F_j(p)$ and $G_j(p)$ are the coefficients in condition (\ref{phiC-phiD>0}). 

In Supplementary Section 5.2, we mathematically prove that inequality (\ref{Hx}) holds for arbitrary $k$, $m$, and $p\in(0,1)$. 
This indicates that, on any homogeneous hypergraph with hyperdegree $k$ and order $m$, and for any fraction of the initial mutant $p\in(0,1)$, imitation-based update rules with higher information diversity promote cooperation more effectively across the entire space of the $m$-player social dilemmas in the sense of the set containment defined above. 
To numerically verify this, instead of calculating $H(k,m,p,x)$ that involve four parameters, we calculate $H_{\text{max}}(k,m,p)=\max_{0 \le x < m-1} H(k,m,p,x)$ that has three parameters and is easier to present. 
This is realized by the fact that $H_{\text{max}}(k,m,p)<0$ is equivalent to inequality (\ref{Hx}).
In Figs. \ref{fig5}e-n, we conduct extensive numerical calculations to examine the sign of $H_{\text{max}}(k,m,p)$ over the following parameter values: $p\in \{0.01,0.02,0.03,0.04,0.05,0.1,0.2,0.3,0.4,0.5\}$ (only the left half is considered due to the symmetry about $p=0.5$) and integer $k,m\in[3,100]$. 
The results show that $H_{\text{max}}(k,m,p) < 0$ for all the parameter values we consider. In summary, we establish that increasing information diversity during the decision-making process universally facilitates the emergence of cooperation, extending from three specific games to general multiplayer social dilemmas. 
\section*{Discussion}
In this work, we propose a class of imitation-based update rules that explicitly incorporate the group structure of higher-order networks, and investigate how they shape the evolution of cooperation. Specifically, when an individual is selected to update its strategy (the focal individual), it randomly selects $s$ hyperedges from all hyperedges it belongs to, and then chooses $q$ neighbors from each selected hyperedge as role models. After this, the focal individual imitates the strategy of one of these role models with a probability proportional to its fitness. In this way, this class of update rules takes into account not only the role models themselves, but also the groups from which they are drawn.
On homogeneous higher-order networks, we derive a closed-form condition for the success of cooperation, which depends on $s$ and $q$ only through the scalar $\eta$. Based on this condition, we uncover a quantity for comparing the effectiveness of different update rules in promoting cooperation, termed information diversity. Information diversity is the probability that among the $sq$ pieces of information acquired by an individual, two randomly selected pieces originate from different groups. We prove that, for arbitrary multiplayer social dilemmas, the relative effectiveness of update rules in promoting cooperation is determined by their information diversity. Extensive simulations show that high information diversity continues to favor cooperation on heterogeneous hypergraphs constructed synthetically and from empirical data. Our work provides a unified perspective to explain the differences in the capability of classical imitation-based update rules to promote cooperation. It identifies information diversity as the decisive feature of update rules that governs the evolution of cooperation on higher-order networks, offering valuable guidance for identifying and designing effective decision-making rules in complex networked systems.
	
On pairwise networks, the death-birth update rule is among the most effective imitation-based update rules for promoting cooperation\cite{ohtsuki2006Nature,wang2023imitation}. It requires individuals to consider information from all their neighbors when updating their strategy. In real-world settings, however, acquiring information entails time, effort, and financial costs. Our study shows that such indiscriminate imitation becomes less effective on higher-order networks. A more efficient way is to select just one neighbor from each hyperedge as a role model, thereby maximizing information diversity. Remarkably, in our model, it suffices for an individual to select two hyperedges to which it belongs and obtain information from one neighbor in each ($s=2, q=1$); this simple rule yields the most favorable outcomes in fostering cooperation with the most economic way (consulting the fewest peers while attaining the maximum information diversity). These findings suggest that, in decision-making processes, the source of information may play a more critical role than the amount of information.
	
	In addition, our work not only yields conclusions that differ from those obtained on pairwise networks, but also offers a deeper understanding of phenomena observed there. A recent study on pairwise networks shows that, when the amount of social information is fixed, incorporating personal information is less effective at promoting cooperation than ignoring it\cite{wang2023imitation}. Moreover, in the absence of personal information, the amount of social information has no effect on evolutionary outcomes. Our work offers a unified perspective for these seemingly inconsistent findings. On pairwise networks, for a given focal individual, any piece of social information and its personal information originate from the same pairwise edge; thus, including personal information reduces informational diversity, thereby inhibiting cooperation. When personal information is ignored, different pieces of social information naturally come from distinct edges, making the amount of social information independent of information diversity. These results underscore the necessity of studying the dynamics of group interactions, as they can refine and extend classical conclusions derived from traditional pairwise networks.
	
From a theoretical perspective, our work contributes in two ways. First, most existing analytical results concern linear interactions; explicit closed-form conditions for nonlinear multiplayer games on structured populations are scarce. Exploring nonlinear higher-order interactions is both theoretically and practically significant, a point recently emphasized by Allen\cite{allen2026understanding}. Our study provides such conditions, expanding the range of multiplayer interaction patterns that can be analyzed. Second, a previous study examines the impact of higher-order network structures on evolutionary outcomes and derives mathematical results for arbitrary hypergraphs\cite{sheng2024strategy}. 
That framework is exact and general in network structure, and this exactness requires numerically solving a system of linear equations of size $O(N^{L+1})$ (where $L$ is the largest group size), which becomes infeasible for large populations and large group sizes. By extending the pair approximation method, we instead obtain explicit closed-form conditions on large homogeneous higher-order networks, trading exactness and structural generality for tractability and interpretability. Such an approach eventually makes it possible for us to answer the important question, namely, what general features of the update rule shape the evolutionary outcomes.
	
Besides, our work indicates that the capabilities of the class of update rules considered here to facilitate cooperation are strictly ordered, in the sense that the corresponding game spaces associated with any two update rules, $\mcal{S}_{\mcal{D}_1}$ and $\mcal{S}_{\mcal{D}_2}$, always exhibit an inclusion relation.
This property is nontrivial.
A previous study\cite{pena2016ordering} comparing the capability of different network structures to promote cooperation defines $\mcal{S}_{\mcal{N}_1}$ ($\mcal{S}_{\mcal{N}_2}$) as the set of games under which cooperation is favored on network structure ${\mcal{N}_1}$ (${\mcal{N}_2}$).
The study showed that the relation between $\mcal{S}_{\mcal{N}_1}$ and $\mcal{S}_{\mcal{N}_2}$ is not necessarily one of inclusion, but may instead involve partial overlap. In such cases, no network structure can be regarded as universally better than another, and the comparison may instead depend on the relative sizes of  $\mcal{S}_{\mcal{N}_1}$ and $\mcal{S}_{\mcal{N}_2}$. The strict ordering we observe may rely on the homogeneity of the structures we consider, which may be a limitation of our analysis. 
Although our simulations indicate that higher information diversity continues to favor cooperation on heterogeneous networks, whether the strict set-containment ordering across the entire game space also holds there remains an open question.
A further limitation concerns the pair approximation method itself\cite{ohtsuki2006Nature}, which assumes that local neighborhoods are approximately tree-like and is therefore not exact on the overlapping, clustered hyperedges of empirical higher-order networks. The agreement between our analytical predictions and simulations on synthetic and empirical hypergraphs suggests that the approach is actually robust, but a rigorous treatment of hyperedge overlap remains an important direction for future work.
	
For other fields, our work may also offer insights. Indeed, our findings provide testable hypotheses for behavioral and social-psychological experiments, where both the number of groups an individual selects and the number of peers sampled within each group can be systematically varied during the decision-making process to assess their effects on prosocial behavior. In addition, our results suggest potential guidelines for organizational and policy design, indicating that access to diverse information sources may promote collective cooperation in complex social systems.  
\section*{Methods}
	Here we briefly outline the mathematical analysis and simulation procedures, with full details provided in the Supplementary Information.
	\subsection*{Theoretical model}
	We consider higher-order networks represented by hypergraphs with $N$ nodes, where each node is
	connected to $k$ hyperedges and each hyperedge contains $m\ge2$ nodes. Denote $p_C$ ($p_D$) as the fraction of $C$-players ($D$-players) on the hypergraph, $p_{iC}$ the fraction of hyperedges that have $i$ $C$-players $(i = 0, 1,..., m)$, and $q_{iC|X}$ is the
	conditional probability that an $X$-player ($X=C$ or $D$) has a hyperedge with $i$ $C$-coplayers $(i=0, 1, ..., m-1)$. Based on these definitions, we have
	$p_{C}+p_{D}=1,~\sum_{i=0}^{m} p_{i C}=1\label{sum-pia},~\sum_{i=0}^{m-1} q_{i C \mid C}=1,~\sum_{i=0}^{m-1} q_{i C \mid D}=1.$ Furthermore, we find that both $p_{iC}$ and $q_{iC|D}$ can be expressed in terms of $p_C$ and $q_{iC|C}$ ($i = 0, 1, \dots, m-1$). Therefore, only $m+1$ state variables are required to fully describe the system dynamics, namely $p_C$ and $q_{iC|C}$ ($i = 0, 1, \dots, m-1$).
	
	Let us first consider the case where a focal $C$-player is randomly chosen to update its strategy.
	Suppose this focal player has a hyperedge that contains $m_C$ $C$-coplayers. Denote $P_{X}^{(m_C)}$ as the fitness of the $X$-coplayer within the hyperedge that contains both the focal $C$-player and its $m_C$ $C$-coplayers. We have $P_{C}^{\left(m_{C}\right)}=\exp \left(w\left((k-1) \sum_{i=0}^{m-1} q_{i C \mid C} a_{i}+a_{m_{C}}\right)/k\right)$,
	and
	$P_{D}^{\left(m_{C}\right)}=\exp \left(w\left((k-1) \sum_{i=0}^{m-1} q_{i C \mid D} b_{i}+b_{m_{C}+1}\right)/k\right),$
	where $w \ge0$ represents the intensity of selection.
	
	Denote $n_{i}$ as the number of hyperedges that contains $i$ $C$-coplayers among the $k$ hyperedges the focal player has, which satisfies $\sum_{i=0}^{m-1}n_i=k$. We denote $s_i$ as the number of hyperedges that have $i$ $C$-coplayers among the $s$ hyperedges, which satisfies $\sum_{i=0}^{m-1} s_{i}=s$. Moreover, we employ $s_i^j$ to denote the number of hyperedges containing $j$ $C$-coplayers among the $q$ individuals chosen from the $s_i$ selected hyperedges. Thus, the probability that the fraction of $C$-players in the population decreases by $1/N$, is
	\begin{eqnarray}\label{-1/N}
		&&\operatorname{Prob}\left(\Delta p_{C}=-\frac{1}{N}\right)
		\nonumber\\
		=&&p_{C} \sum_{n_{0}+\cdots=k} \frac{k !}{n_{0} ! \cdots n_{m-1} !} q_{0 C \mid C}^{n_{0}} \cdots q_{(m-1) C \mid C}^{n_{m-1}}\sum_{s_{0}+\cdots=s}\frac{\prod_{ l=0}^{m-1}\binom{n_l}{s_l}}{\binom{k}{s}}\sum_{s^0_{0}=s_0}\cdots \sum_{s^{m-1}_{0}+\cdots+s^{m-1}_{m-1}=s_{m-1}}\nonumber\\
		&&\prod_{i=0}^{m-1}\frac{s_i!}{s^i_0!s^i_1!\cdots s^i_{i}!}\prod_{j=0}^i\left[ \frac{\binom{i}{j}\binom{m-1-i}{q-j}}{\binom{m-1}{q}}\right]^{s^i_j}\frac{\sum_{i=0}^{m-1}\sum_{j=0}^{i}s^i_j(q-j)P_{D}^{(i)}}{\sum_{i=0}^{m-1}\sum_{j=0}^{i} s^i_jjP_{C}^{(i)}+\sum_{i=0}^{m-1}\sum_{j=0}^{i}s^i_j(q-j)P_{D}^{(i)} }.\nonumber\\
	\end{eqnarray}
	Meanwhile, the probability that the fraction of hyperedges with $j$ $C$-players ($1\le j \le m-1$) has a net change of $\Delta p_{j C}=-\frac{m n_{j-1}}{N k}+\frac{m n_{j}}{N k}=\frac{m\left(n_{j}-n_{j-1}\right)}{N k}$ is
	\begin{eqnarray}\label{-pja}
		&\quad&\operatorname{Prob}\left(\Delta p_{j C}=\frac{m\left(n_{j}-n_{j-1}\right)}{N k}\right)\nonumber\\
		&=&p_{C} \sum_{s_{0}+\cdots=s}\frac{\prod_{ l=0}^{m-1}\binom{n_l}{s_l}}{\binom{k}{s}}\sum_{s^0_{0}=s_0}\cdots \sum_{s^{m-1}_{0}+\cdots+s^{m-1}_{m-1}=s_{m-1}}\prod_{i=0}^{m-1}\frac{s_i!}{s^i_0!s^i_1!\cdots s^i_{i}!}\prod_{j=0}^i\left[ \frac{\binom{i}{j}\binom{m-1-i}{q-j}}{\binom{m-1}{q}}\right]^{s^i_j}\nonumber\\
		&&\frac{\sum_{i=0}^{m-1}\sum_{j=0}^{i}s^i_j(q-j)P_{D}^{(i)}}{\sum_{i=0}^{m-1}\sum_{j=0}^{i} s^i_jjP_{C}^{(i)}+\sum_{i=0}^{m-1}\sum_{j=0}^{i}s^i_j(q-j)P_{D}^{(i)} }\nonumber\\
		&=&p_C-\frac{p_C}{k(m-1)}\sum_{i=0}^{m-1}n_ii+O(w).
	\end{eqnarray}
	For $j=0$ or $j=m$, in a similar way, we have 
	\begin{equation}
		\operatorname{Prob}\left(\Delta p_{0C}=\frac{m n_{0}}{N k}\right)=
		\operatorname{Prob}\left(\Delta p_{m C}=-\frac{m n_{m-1}}{N k}\right)	p_C-\frac{p_C}{k(m-1)}\sum_{i=0}^{m-1}n_ii+O(w).
	\end{equation}
	
	Next, we consider the case where a $D$-player is chosen as the focal individual and imitates one of
	its $C$-coplayers. Similarly, we can derive the probability that an $C$-coplayer replaces the focal $D$-player,  namely, $\operatorname{Prob}\left(\Delta p_{C}=\frac{1}{N}\right)$. We can also obtain the probabilities of net changes in the fraction of hyperedges with $j$ $C$-players, namely, $\operatorname{Prob}\left(\Delta p_{mC} = \frac{m n_{m-1}}{N k}\right)$, $\operatorname{Prob}\left(\Delta p_{0C} = -\frac{m n_{0}}{N k}\right)$, and $\operatorname{Prob}\left(\Delta p_{jC} = \frac{m(n_{j-1} - n_{j})}{N k}\right)$ for $1 \le j \le m-1$. 
	
	Based on the above results, we now derive the rate of change in the fraction of $C$-player, $\dot{p}_{C}$, and the rate of change of the fraction of hyperedges containing $j$ $C$-players, $\dot{p}_{jC}$. First, $\dot{p}_{C}$ can be written as
	\begin{equation}\label{D(pC)}
		\dot{p}_{C}=\frac{1}{N} \operatorname{Prob}\left(\Delta p_{C}=\frac{1}{N}\right)+\left(- \frac{1}{N}\right)  \operatorname{Prob}\left(\Delta p_{C}=-\frac{1}{N}\right).
	\end{equation}
	For $1 \leq j \leq m-1$, the derivative of $p_{j C}$ can be expressed as 
	\begin{eqnarray}\label{Dpja}
		\dot{p}_{j C}&=&p_D  \sum_{n_{0}+\cdots=k} \frac{k !}{n_{0} ! \cdots n_{m-1} !} q_{0C \mid D}^{n_{0}} \cdots q_{(m-1) C \mid D}^{n_{m-1}}\frac{m\left(n_{j-1}-n_{j}\right)}{N k} \operatorname{Prob}\left(\Delta p_{j C}=\frac{m\left(n_{j-1}-n_{j}\right)}{N k}\right) \nonumber\\
		&&-p_C \sum_{n_{0}+\cdots=k} \frac{k !}{n_{0} ! \cdots n_{m-1} !} q_{0 C \mid C}^{n_{0}}  \cdots q_{(m-1) C \mid C}^{n_{m-1}}\frac{m\left(n_{j-1}-n_{j}\right)}{N k} \operatorname{Prob}\left(\Delta p_{j C}=\frac{m\left(n_{j}-n_{j-1}\right)}{N k}\right). \nonumber \\
	\end{eqnarray}
	Similarly, we can derive the $\dot{p}_{0 C}$ and $\dot{p}_{m C}$. 
	
	Furthermore, $\dot q_{iC|C}$ can be expressed in terms of $\dot p_{(i+1)C}$, $p_{(i+1)C}$, and $p_C$, resulting in $m$ equations for $\dot q_{iC|C}$ ($i = 0, 1, \dots, m-1$). Together with equation (\ref{D(pC)}), this set of equations form a closed dynamical system that describes the evolutionary dynamics on higher-order networks with $m+1$ state variables. Under weak selection $0<w\ll 1$, by applying the time scale separation, the dynamical system admits a slow manifold, which leads to 
	\begin{equation}\label{slowmanifold1}
		\sum_{i=0}^{m-1} i\left(q_{iC|C} - q_{iC|D}\right) = \frac{1}{k - 1}.
	\end{equation}
	By using equation (\ref{slowmanifold1}), we can further express $q_{iC|D}$ and $q_{iC|C}$ as polynomial functions of $p_C$, denoted by $q_{iC|D}(p_C)$ and $q_{iC|C}(p_C)$, respectively. Thus, the variables describing the system can be simplified from $p_C$ and $q_{iC|C}$ ($i = 0, 1, \dots, m-1$) to only $p_C$.
	
	To calculate the fixation probability of strategy $C$ and $D$, i.e., $\phi_C$ and $\phi_D$, we employ the backward Kolmogorov equation \cite{ohtsuki2006Nature}. Specifically, given the initial condition $p_C(t=0)=y$, the backward Kolmogorov equation is
	\begin{equation}\label{Kolmogorov}
		M(y)\frac{d\phi_{C}(y)}{dy}+\frac{V(y)}{2}\frac{d^2\phi_C(y)}{dy^2}=0.
	\end{equation}
	where $M(y)$ and $V(y)$ are related to the the expectation and variance of $\Delta p_{C}$, respectively. And the boundary conditions are given by
	$\phi_{C}(1)=1 ~ \text{and} ~\phi_{C}(0)=0$.
	The expectation and variance of $\Delta p_C$ over a short time interval $\Delta t$ are denoted by $E[\Delta p_C]$ and $\mathrm{Var}[\Delta p_C]$, respectively. We have $M(\Delta p_C)=E[\Delta p_C]/\Delta t$ and $V(\Delta p_C)=V[\Delta p_C]/\Delta t$. 
	Within the short time interval, the expectation and variance of $\Delta p_C$ can be written as
	\begin{equation}
		M\left[\Delta p_{C}\right]\approx\dot p_C
	\end{equation}
	and
	\begin{equation}
		\text{Var}[\Delta p_{C}]\approx \left(\frac{1}{N} - E[\Delta p_{C}]\right)^{2} \operatorname{Prob} \left(\Delta p_{C}=\frac{1}{N} \right) 
		+\left(-\frac{1}{N} - E[\Delta p_{C}]\right)^{2} \operatorname{Prob}\left(\Delta p_{C}=-\frac{1}{N} \right). 
	\end{equation}
	By solving equation (\ref{Kolmogorov}), we can obtain the fixation probability of strategy $C$. Furthermore, from  the definition of fixation probabilities, we can get $\phi_C(p)+\phi_D(1-p)=1$. Thus, we can also derive the fixation probability of strategy $D$. 

When $(k-1)(m-1)>1$ and $sq\ne 1$, we have that strategy $C$ is favored over strategy $D$ if $\phi_C(p)-\phi_D(p)>0 $, which implies equation (\ref{phiC-phiD>0}) in the main text.  For $F_j(p)$ and $G_j(p)$ in equation (\ref{phiC-phiD>0}), we have
	\begin{equation}\label{F}
		F_j(p)=tI_{j,1}^p,~ 0\le j\le m-1
	\end{equation}
	and
	
	\begin{equation}\label{G}
		G_j(p) = 
		\left\{
		\begin{array}{ll}
			(m-1-j)I_{j,0}^p, & \hbox{$0\le j<m-1$;} \\
			0, & \hbox{$j=m-1$,}
		\end{array}
		\right.
	\end{equation}
	where
	\begin{eqnarray}
		I_{j,z}^p &=& \binom{m-1}{j} \sum_{l=0}^j \sum_{i=1-z}^{m-1-j} \sum_{g=0}^l \binom{l}{g}u(m-1-j, i) u(j,l)  \left[ 1-p^{g+i+z+1}-(1-p)^{g+i+z+1}\right] \nonumber\\
		& &\frac{(-1)^g t^{g+i-1}(t+1)^{l-g}}{(g+i+z)(g+i+z+1)}.
	\end{eqnarray}	
	Here, $u(j,l)$ denotes the unsigned Stirling number of the first kind, satisfing the recurrence relation $u(n+1,l)=u(n,l)n+u(n,l-1)$ for $l>0$ with $u(0,0)=1,~u(n,0)=u(0,n)=0$ for $n>0$. 
In the meanwhile, for interpretation, we can rewrite $F_j(p)$ and $G_j(p)$ as
	\begin{equation}
		F_j(p)=\frac{(t+m-1)!}{t!}\int_0^{p} \int_v^{1-v}q_{jC|C}(u)dudv,
	\end{equation}
and 
	\begin{equation}
		G_j(p) = 
		\frac{(t+m-1)!}{(k-1)t!}\int_0^{p}\!\!\int_v^{1-v}q_{jC|CD}(u)~ du dv,
	\end{equation}
when $j<m-1$ and $G_{m-1}(p)=0$ (see Supplementary Section 5.4 for detailed derivations). Here, $q_{jC|CD}$ ($j=0,1,\cdots,m-2$) is the conditional probability that a hyperedge contains $j$ additional $C$ individuals, given that an $C-D$ pair is observed in that hyperedge.
\subsection*{Simulation process}
As an illustrative example, Algorithm 1 summarizes the simulation procedure used to compute the fixation probability of cooperators $\phi_C$ in the LPGG. For each value of the synergy factor $r_1$, we perform $Z = 10^7$ independent replicates. 
At the beginning of the $z$th replicate, $pN$ individuals are randomly initialized as cooperators. In each round, every individual participates in the multiplayer game within all hyperedges to which it belongs. 
The payoff of individual $i$ is computed as the average payoff obtained from all hyperedges in which it participates. Next, one individual is randomly selected as the focal individual for strategy updating. 
The focal individual randomly samples $s$ hyperedges from its hyperedges and selects $q$ neighbors from each sampled hyperedge as role models. 
It then adopts the strategy of one role model $j$ with a probability proportional to $j$'s fitness. The process is iterated until the population reaches one of the two absorbing states, namely full cooperation or full defection (i.e., the number of cooperators $n_C$ equals to $N$ or 0). Let $z_C$ denote the number of replicates that end in full cooperation. 
The fixation probability of cooperators is defined as $\phi_C = z_C/Z$.

\begin{algorithm}[htbp]
    \floatname{algorithm}{Algorithm}
	\caption{Simulation procedure for computing the fixation probability of cooperators in the LPGG}
	\begin{algorithmic}[1]
		
		\For{each synergy factor $r_1$}
		
		\State $z_C \gets 0$ 
		
		\For{$z = 1,\dots,Z$}
		
		\State Randomly initialize $pN$ cooperators
		\State $n_C \gets pN$
		
		\Repeat
		
		\For{each individual $i = 1,\dots,N$}
		\State $\pi_i \gets 0$
		\For{each hyperedge containing $i$}
		\State Compute the number of co-cooperators $j$ in the hyperedge
		
		\If{$i$ is a cooperator}
		\State $\pi_i \gets \pi_i+\frac{(j+1)r_1c}{m}-c$
		\Else
		\State$\pi_i \gets \pi_i+\frac{jr_1c}{m}$
		\EndIf

		\EndFor

		\State $\pi_i \gets \pi_i/k_i$ 
		\EndFor
		
		\State Randomly select a focal individual $l$
		\State Randomly sample $s$ hyperedges from its $k_l$ hyperedges
		
		\State $\Omega_l^{(s,q)} \gets \emptyset$
		\For{each sampled hyperedge}
		\State Randomly select $q$ neighbors as role models and add them to $\Omega_l^{(s,q)}$
		\EndFor
		
		\State Individual $l$ imitates $j \in \Omega_l^{(s,q)}$ with probability
		\[
		p^{(s,q)}_{l \to j}
		=
		\frac{\exp(w \pi_j)}
		{\sum_{x \in \Omega_l^{(s,q)}} \exp(w \pi_x)}
		\]
		
		\State Update $n_C$
		
		\Until{$n_C = 0$ or $n_C = N$}
		
		\If{$n_C = N$}
		\State $z_C \gets z_C + 1$
		\EndIf
		
		\EndFor
		
		\State $\phi_C(r_1) \gets z_C / Z$
		
		\EndFor
		
	\end{algorithmic}
\end{algorithm}

\clearpage
\bibliographystyle{naturemag}

\clearpage

\makeatletter
\@fpsep\textheight
\makeatother

\clearpage

\clearpage

\clearpage

\clearpage

\end{document}